\documentclass[twocolumn,showpacs,preprintnumbers,amsmath,amssymb]{revtex4}
\usepackage{amsfonts}
\usepackage{mathrsfs}
\usepackage{graphicx}% Include figure files
\usepackage{dcolumn}% Align table columns on decimal point
\usepackage{bm}% bold math

\begin{document}

\title{The Decay of Multiqudit Entanglement}

\author{Zhao Liu}%
 \email{liuzhaophys@aphy.iphy.ac.cn}
\author{Heng Fan}
 \email{hfan@aphy.iphy.ac.cn}
\affiliation{%
Institute of Physics, Chinese Academy of Sciences, Beijing 100190,
China
}%
\date{\today}% It is always \today, today,

\begin{abstract}
We investigate the decay of entanglement of a generalized $N$-qudit
GHZ state
%$|\Psi_{d}\rangle=\sum_{i=0}^{d-1}\alpha_{i}|i\rangle^{\otimes N}$,
with each qudit passing through independently in a quantum noisy
channel. By studying the time at which the entanglement completely
vanishes and the time at which the entanglement becomes arbitrarily
small, we try to find how the robustness of entanglement is
influenced by dimension $d$ and the number of particles $N$.
\end{abstract}

\maketitle

%\section{\label{sec:level1}First-level heading:\protect\\ The line
%break was forced \lowercase{via} \textbackslash\textbackslash}

{\label{sec:level1}} \section{Introduction} Quantum entanglement, as
the most non-classical phenomenon in quantum mechanics, lies in the
central position of quantum information theory and has been
identified as a key resource in many applications such as quantum
teleportation, quantum key distribution and quantum computation
\cite{r1}, see Ref.\cite{h} for a review. For large-scale quantum
information processing, multiparticle entanglement is indispensable.
Therefore the understanding of the dynamical property of
multiparticle entanglement in realistic environment is of
fundamentally importance.

Due to the interaction with environment, multiparticle entanglement
decays inevitably. In the past several years, there are many
excellent papers concerning with the robustness of multiparticle
entanglement under the influence of environment
\cite{r2,r3,r7,r8,r9,r4,r11,r10}. It was shown that for a $N$-qubit
GHZ state (throughout $N$ is even for simplicity)
$\frac{1}{\sqrt{2}}(|0\rangle^{\otimes N}+|1\rangle^{\otimes N})$
with each qubit exposed to a depolarizing channel independently, the
bipartite entanglement of $(N-n)|n$ bipartition disappears in finite
time which is named as entanglement sudden death (ESD) in Refs.
\cite{r2,r8,r11,r10}. The entanglement which first disappears is
that corresponding to $1|(N-1)$ bipartition and the time at which
this entanglement disappears decreases with $N$. The entanglement
which last disappears is that corresponding to $N/2|N/2$ bipartition
and the time at which this entanglement disappears grows with $N$
\cite{r2,r3}. Aolita \emph{et al}. \cite{r4} proposed that the time
at which the entanglement becomes arbitrarily small is a better
quantity characterizing the robustness of entanglement than the ESD
time and they found that this time is inversely proportional to $N$
for $N$-qubit generalized GHZ state
$|\Psi_{2}\rangle=\alpha|0\rangle^{\otimes
N}+\beta|1\rangle^{\otimes N}$.

It is understandable that the robustness of multiparticle
entanglement not only depends on the number of particles but also on
the dimensionality of each particle's Hilbert space. And for
different quantum channels, the behavior of the multiparticle
entanglement will also be different. However, it is necessary to
have a fixed example to see how all of these results can be
obtained. In this article, we will study those problems. We suppose
each particle is in a $d$-dimensional Hilbert space and it is called
generally a qudit. We consider a simple case of a generalized
$N$-qudit GHZ state with each qudit interacts with a quantum channel
independently and find how does the robustness of entanglement
change with the increase of $d$ or $N$ through studying the time
evolution of the bipartite entanglement of $(N-n)|n$ bipartition. We
use the depolarizing channel and the phase damping channel as the
quantum channels.

{\label{sec:level1}} \section{Quantum Channels} First we would like
to introduce two $d\times d$ matrices extremely useful in
constructing our quantum channels. They are defined as
$X|i\rangle=|i+1\rangle$(mod $d$) and
$Z|i\rangle=\omega^{i}|i\rangle$, where $i=0,...,d-1$ and
$\omega=\exp(2\pi\textrm{i}/d)$. One can easily see that when $d=2$,
they are just Pauli-sigma $x$ and Pauli-sigma $z$ matrix
respectively. Now we will construct three operator transformations
through $X$ and $Z$:
\begin{eqnarray}
\mathcal {E}_{1}(\mathcal {A})=(1-p)\mathcal
{A}+\frac{p}{d^{2}}\sum_{ij=0}^{d-1}X^{i}Z^{j}\mathcal {A}Z^{\dagger
j}X^{\dagger i},\nonumber\\
\mathcal {E}_{2}(\mathcal {A})=(1-p)\mathcal
{A}+\frac{p}{d}\sum_{i=0}^{d-1}Z^{i}\mathcal {A}Z^{\dagger i},
\label{e1}
\end{eqnarray}
where $p\in[0,1]$ and $\mathcal {A}$ is an arbitrary operator on
$d$-dimensional Hilbert spaces. When $\mathcal {A}$ is a density
matrix, we can see $\mathcal {E}_{i},i=1,2$ as three quantum
channels. Through calculation, it can be shown that for an input
density matrix $\rho$, $\mathcal
{E}_{1}(\rho)=(1-p)\rho+\frac{p}{d}\mathbf{1}$ and $\mathcal
{E}_{2}(\rho)=(1-p)\rho+p\sum_{k=0}^{d-1}\rho_{kk}|k\rangle\langle
k|$, where $\mathbf{1}$ is a $d\times d$ identity matrix. It's
obvious to see that in fact $\mathcal {E}_{1}$ is a depolarizing
channel and $\mathcal {E}_{2}$ is a phase damping channel.

{\label{sec:level2}} \section{The evolution of entanglement} A
generalized $N$-qudit GHZ state can be written as
$|\Psi_{d}\rangle=\sum_{i=0}^{d-1}\alpha_{i}|i\rangle^{\otimes N}$,
where $\alpha_{i}$ is a complex number and
$\sum_{i=0}^{d-1}|\alpha_{i}|^{2}=1$. Here we want to show how does
the bipartite entanglement corresponding to $(N-n)|n$ bipartition of
it evolve under the influence of the three quantum channels
constructed above. In this article we adopt negativity as the
measure of entanglement \cite{r5}.

When each qudit is exposed to a depolarizing channel, we can
calculate $\Big(\bigotimes_{i=1}^{N}\mathcal
{E}_{1,i}\Big)(|\Psi_{d}\rangle\langle\Psi_{d}|)$ and denote it as
$\rho_{1}(p)\equiv\sum_{i=0}^{d-1}\sum_{k=0}^{N}\sum_{\mathcal
{P}}|\alpha_{i}|^{2}\Big(\frac{p}{d}\Big)^{k}\Big(1-\frac{d-1}{d}p\Big)^{N-k}
\times\mathcal {P}\Big[(|i\rangle\langle
i|)^{\otimes(N-k)}\otimes\Big(\sum_{j=0,j\neq
i}^{d-1}|j\rangle\langle j|\Big)^{\otimes k}\Big]
+(1-p)^{N}\sum_{i=0}^{d-1}\sum_{j=0,j\neq
i}^{d-1}\alpha_{i}\alpha_{j}^{*}(|i\rangle\langle j|)^{\otimes N},$
where $\mathcal {P}$ means all possible permutations. Partial
transposing the part of $n$ particles of $\rho_{1}(p)$ and noting
that the first term of $\rho_{1}(p)$ is diagonal so it will not be
changed after partial transpose, we have
$\rho_{1}(p)^{\Gamma_{n}}=\sum_{i=0}^{d-1}\sum_{k=0}^{N}\sum_{\mathcal
{P}}|\alpha_{i}|^{2}\Big(\frac{p}{d}\Big)^{k}\Big(1-\frac{d-1}{d}p\Big)^{N-k}
\times\mathcal {P}\Big[(|i\rangle\langle i|)^{\otimes
(N-k)}\otimes\Big(\sum_{j=0,j\neq i}^{d-1}|j\rangle\langle
j|\Big)^{\otimes k}\Big] +(1-p)^{N}\sum_{i=0}^{d-1}\sum_{j=0,j\neq
i}^{d-1}\alpha_{i}\alpha_{j}^{*}(|i\rangle\langle
j|)^{\otimes(N-n)}\otimes(|j\rangle\langle i|)^{\otimes n}$. There
are $\frac{d(d-1)}{2}$ eigenvalues $\mu$ of
$\rho_{1}(p)^{\Gamma_{n}}$ that can be negative and are determined
by the smaller eigenvalues of the following $\frac{d(d-1)}{2}$
$2\times2$ matrices $ \left(
\begin{array}{cccc}
\lambda_{n}^{ij}&\alpha_{i}\alpha_{j}^{*}(1-p)^{N}\\
\alpha_{i}^{*}\alpha_{j}(1-p)^{N}&\lambda_{N-n}^{ij}
\end{array}\right),
\label{e4}
$
where $i<j$ and
$\lambda_{n}^{ij}=|\alpha_{i}|^{2}\Big(\frac{p}{d}\Big)^{n}\Big(1-\frac{d-1}{d}p\Big)^{N-n}
+|\alpha_{j}|^{2}\Big(\frac{p}{d}\Big)^{N-n}\Big(1-\frac{d-1}{d}p\Big)^{n}$.
One can easily derive
$\mu_{n}^{ij}=\xi_{n}^{ij}-\sqrt{(\xi_{n}^{ij})^{2}
-\eta_{n}^{ij}}$, where
$\xi_{n}^{ij}=\frac{1}{2}(\lambda_{n}^{ij}+\lambda_{N-n}^{ij})$ and
$\eta_{n}^{ij}=\lambda_{n}^{ij}\lambda_{N-n}^{ij}-|\alpha_{i}\alpha_{j}|^{2}(1-p)^{2N}$.
Then we can define $\mathcal {N}_{n}^{ij}=\max\{-\mu_{n}^{ij},0\}$
and the negativity can be obtained by summation $\mathcal
{N}_{n}(\rho_{1}(p))=\sum_{i<j=0}^{d-1}\mathcal {N}_{n}^{ij}$.

Similar to the case of qubit \cite{r4}, $\mathcal
{N}_{1}^{ij}\leq\mathcal {N}_{2}^{ij}\leq...\leq\mathcal
{N}_{N/2}^{ij}$ for a given pair of $\alpha_{i}$ and $\alpha_{j}$,
which straightforwardly leads to $\mathcal {N}_{1}\leq\mathcal
{N}_{2}\leq...\leq\mathcal {N}_{N/2}$. So the bipartite entanglement
corresponding to the most balanced partition still disappears last
whereas the one corresponding to the least balanced partition
disappears first. Let $p_{n}=\max_{i<j}p_{n}^{ij}$, where
$p_{n}^{ij}$ is the solution of the equation $\mu_{n}^{ij}=0$, then
$\mathcal {N}_{n}(\rho_{1}(p_{n}))=0$.
%\begin{figure}
%\includegraphics[height=8cm,width=\linewidth]{Graph1.eps}
%\caption{\label{fig:epsart} We choose $N=8$ and $d=3$ to show the
%change of $\mathcal {N}_{n}^{ij}$ with $n$ at different values of
%$p$. Here $\alpha_{i}=\sqrt{8/9}$ and $\alpha_{j}=\sqrt{1/9}$.}
%\end{figure}
Now we investigate dynamical property of $\mathcal {N}_{N/2}$, which
disappears last. First we want to know when does it completely
vanish. By solving $\mu_{N/2}^{ij}=0$, it's easy to find that
\begin{widetext}
\begin{eqnarray}
p_{N/2}^{ij}=\frac{2|\alpha_{i}\alpha_{j}|^{\frac{2}{N}}d}{2|\alpha_{i}\alpha_{j}|^{\frac{2}{N}}d+(|\alpha_{i}|^{2}+|\alpha_{j}|^{2})^{\frac{1}{N}}
\Big\{(|\alpha_{i}|^{2}+|\alpha_{j}|^{2})^{\frac{1}{N}}+\sqrt{4|\alpha_{i}\alpha_{j}|^{\frac{2}{N}}+(|\alpha_{i}|^{2}+|\alpha_{j}|^{2})
^{\frac{2}{N}}}\Big\}},
\label{e5}
\end{eqnarray}
\end{widetext}
which coincides with Eq.(6) of Ref.\cite{r4} when $d=2$. After $p$
reaches the value $p_{N/2}=\max_{i<j}p_{N/2}^{ij}$, $\mathcal
{N}_{N/2}=0$. It's obvious that $p_{N/2}^{ij}<1$ so $p_{N/2}<1$ and
so ESD happens. Before $p=p_{N/2}$, $\mathcal {N}_{N/2}>0$ and
$\rho_{1}(p)$ must be still entangled. Second we want to know when
does $\mathcal {N}_{N/2}$ become arbitrarily small, which is
practically important because before the entanglement is zero, it
can be so small that it's useless as a resource. Like Ref.\cite{r4},
we suppose $\epsilon$ is an arbitrarily small positive number and
define a critical probability $p_{\epsilon}^{ij}$ such that
$\mu_{N/2}^{ij}(p_{\epsilon}^{ij})=\epsilon\mu_{N/2}^{ij}(0)$, which
leads to an equation
\begin{eqnarray}
(|\alpha_{i}|^{2}+|\alpha_{j}|^{2})\Big(\frac{p_{\epsilon}^{ij}}{d}\Big)^{\frac{N}{2}}\Big(1-\frac{d-1}{d}p_{\epsilon}^{ij}\Big)^{\frac{N}{2}}
\nonumber\\
-|\alpha_{i}\alpha_{j}|(1-p_{\epsilon}^{ij})^{N}=-\epsilon|\alpha_{i}\alpha_{j}|.
\label{e6}
\end{eqnarray}
When $p\geq p_{\epsilon}\equiv\max_{i<j}p_{\epsilon}^{ij}$, we can
think $\mathcal {N}_{N/2}$ is too small to be used as a resource.

As the second case let's consider the situation where each qudit is
exposed to a phase damping channel. Similar to the case of
depolarizing channel, it can be obtained that
$\rho_{2}(p)\equiv\Big(\bigotimes_{i=1}^{N}\mathcal
{E}_{2,i}\Big)(|\Psi_{d}\rangle\langle\Psi_{d}|)=\sum_{i=0}^{d-1}|\alpha_{i}|^{2}(|i\rangle\langle
i|)^{\otimes N}+(1-p)^{N}\sum_{i=0}^{d-1}\sum_{j=0,j\neq
i}^{d-1}\alpha_{i}\alpha_{j}^{*}(|i\rangle\langle j|)^{\otimes N}$
and the partial transpose of it is
$\rho_{2}(p)^{\Gamma_{n}}=\sum_{i=0}^{d-1}|\alpha_{i}|^{2}(|i\rangle\langle
i|)^{\otimes N}+(1-p)^{N}\sum_{i=0}^{d-1}\sum_{j=0,j\neq
i}^{d-1}\alpha_{i}\alpha_{j}^{*}(|i\rangle\langle
j|)^{\otimes(N-n)}\otimes(|j\rangle\langle i|)^{\otimes n}$. So it's
easy to verify that the negativity $\mathcal {N}_{n}(\rho_{2}(p))$
can be expressed as $\sum_{i<j=0}^{d-1}\max\{0,-\nu_{n}^{ij}\}$,
where $\nu_{n}^{ij}=-|\alpha_{i}\alpha_{j}|(1-p)^{N}$ is independent
of $n$. We note that for any $n$, $\mathcal {N}_{n}(\rho_{2}(p))=0$
only when $p=1$, meaning that no ESD happens for phase damping
channel. When the negativity becomes arbitrarily small, namely
$\nu_{n}^{ij}(p_{\epsilon}^{ij})=\epsilon\nu_{n}^{ij}(0)$, we have
an equation $(1-p_{\epsilon}^{ij})^N=\epsilon$.

In what follows we will focus our attention on the results of
entanglement evolution under the depolarizing channel and phase
damping channel to study how does the robustness of entanglement
change with $N$ and $d$.

{\label{sec:level3}} \section{Robustness of entanglement} First we
fix $d$ to study the relation between the entanglement robustness
and $N$. For the depolarizing channel, Eq.(\ref{e5}) tells us that
the ESD time of $\mathcal {N}_{N/2}$ grows with $N$ (FIG.1). Noting
that $\lim _{N\rightarrow\infty}p_{N/2}^{ij}=2d/(2d+1+\sqrt{5})$
which is independent of $i$ and $j$, we find
$p_{N/2}=\max_{i<j}p_{N/2}^{ij}$ becomes closer to this value while
$N$ grows (FIG 1). Moreover, when $N$ is big enough, considering
$p_{\epsilon}^{ij}$ is very small, the first term in the LHS of
Eq.(\ref{e6}) can be omitted, leading to
$p_{\epsilon}\sim-(1/N)\ln\epsilon$ (FIG.2). For the phase damping
channel, no matter what $N$ is, no ESD happens. However,
$p_{\epsilon}\sim-(1/N)\ln\epsilon$ still holds. We can see that so
long as $d$ is fixed, the scaling relation between $p_{\epsilon}$
and $N$ is always the same with that in Ref.\cite{r4}, where $d=2$.

Then we fix $N$ to study the relation between the entanglement
robustness and $d$. For the depolarizing channel, from
Eqs.(\ref{e5}) and (\ref{e6}) one can find that when the
coefficients $\alpha_{i}$s are given (it's not necessary to set all
$\alpha_{i}$s the same), both $p_{N/2}^{ij}$ (also $p_{N/2}$) and
$p_{\epsilon}^{ij}$ (also $p_{\epsilon}$) grow with $d$, meaning the
entanglement is more robust (FIG.1 and 2). Moreover, when $d$ is
large enough, we have $\lim_{d\rightarrow\infty}p_{N/2}=1$ (FIG.1)
and $\lim_{d\rightarrow\infty}p_{\epsilon}=1-\epsilon^{1/N}$
(FIG.2). For the phase damping channel, when $N$ is fixed, it's
obvious that both the ESD time (infinity) and
$p_{\epsilon}^{ij}=1-\epsilon^{1/N}$ are independent of $d$.
\begin{figure}
\includegraphics[height=9cm,width=\linewidth]{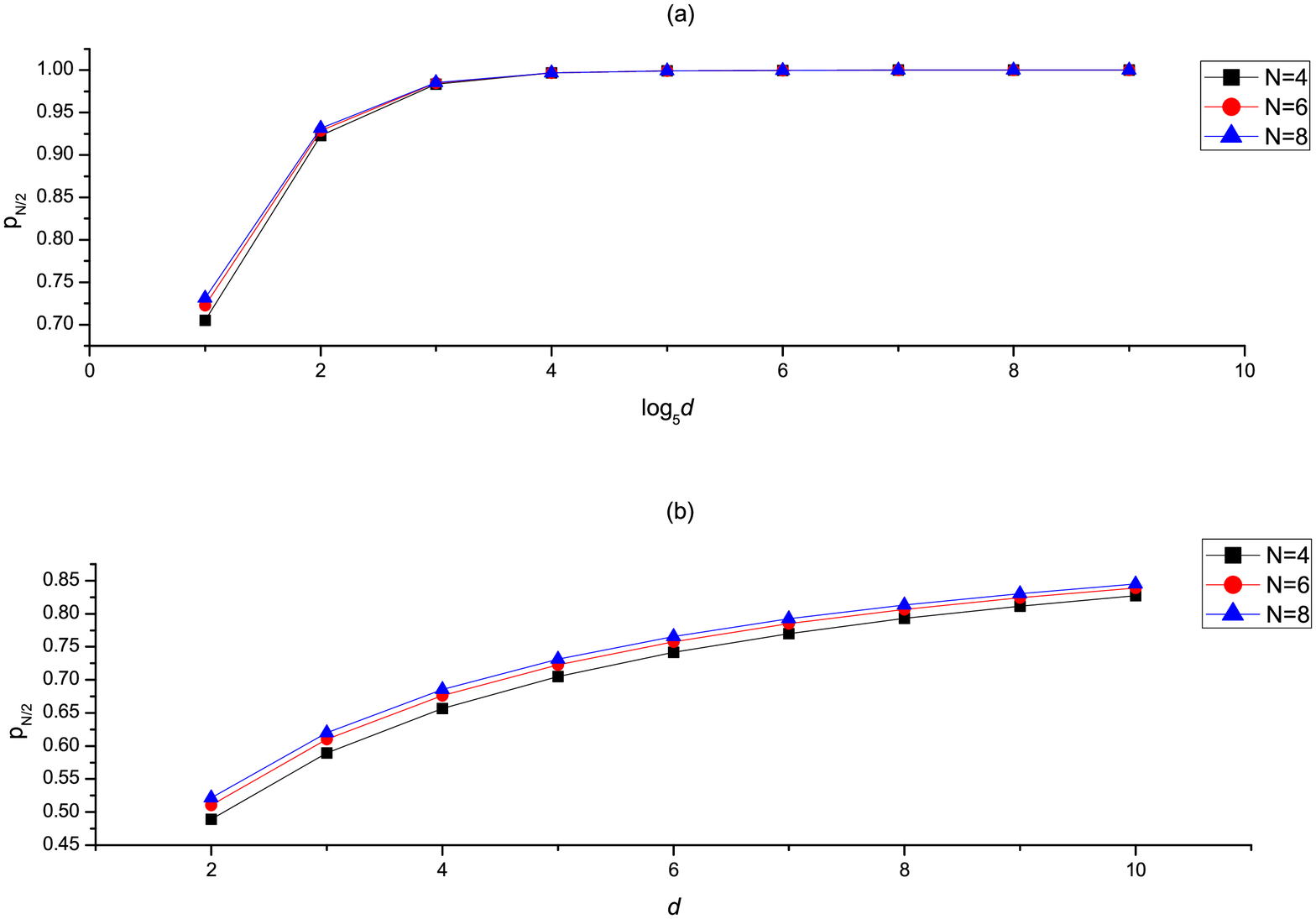}
\caption{\label{fig:epsart} (color online) For the depolarizing
channel, we demonstrate how can $p_{N/2}$ be influenced by both $N$
and $d$. Here we choose $\alpha_{i}=1/\sqrt{d}$ for convenience and
$N=4$ (black cubic), $N=6$ (red circle) and $N=8$ (blue triangle)
respectively. (a): the behavior of $p_{N/2}$ for large $d$. (b): the
behavior of $p_{N/2}$ for small $d$. It's easy to see that $p_{N/2}$
grows with $N$ for a fixed value of $d$ saturating to
$2d/(2d+1+\sqrt{5})$ and it also grows with $d$ and saturates to 1
for any fixed $N$.}
\end{figure}

\begin{figure}
\includegraphics[height=9cm,width=\linewidth]{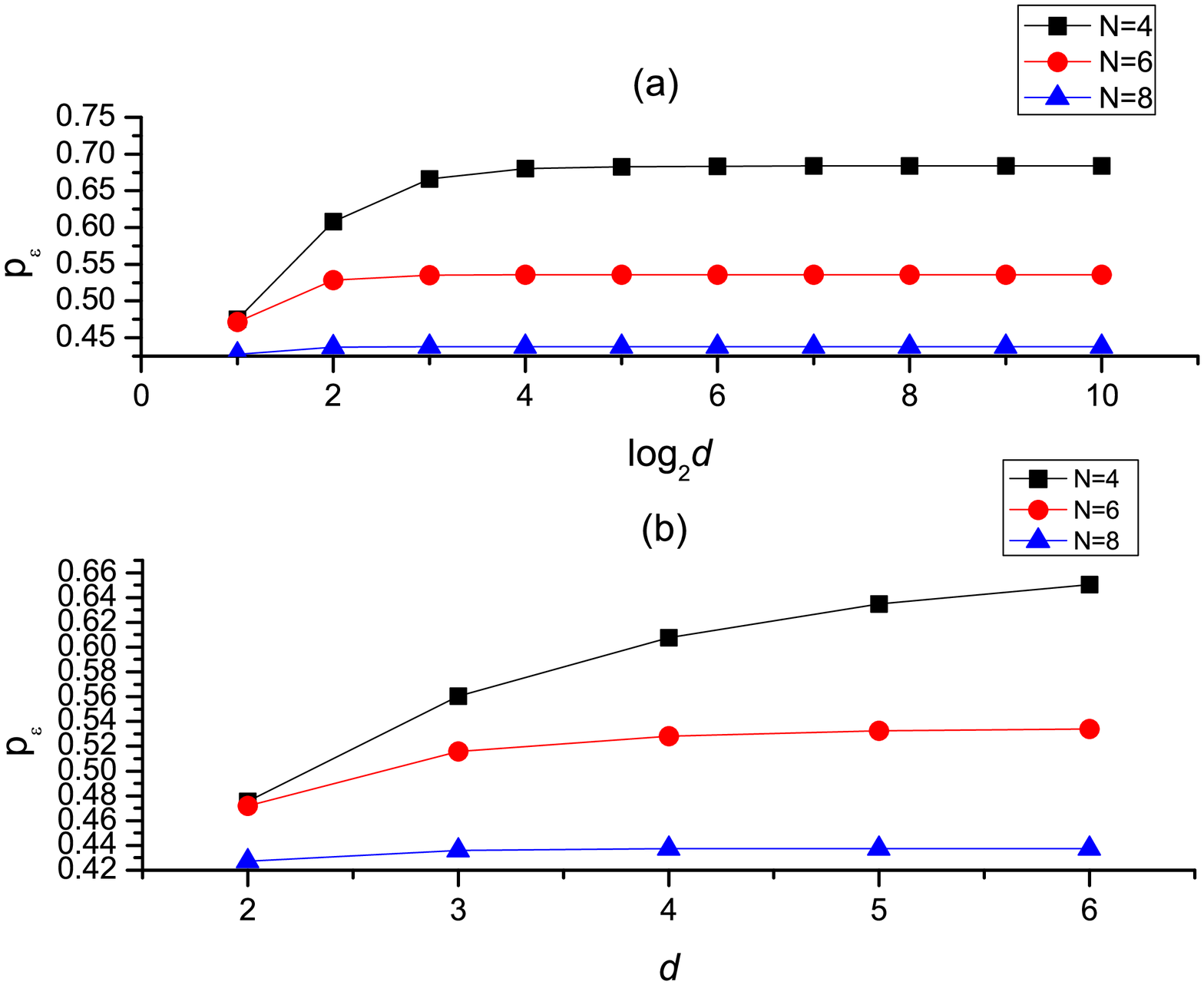}
\caption{\label{fig:2} (color online) For the depolarizing channel,
we show the relation of $p_{\epsilon}$ with $N$ and $d$. Here we
choose $\alpha_{i}=1/\sqrt{d}$ for convenience, $\epsilon=0.01$ and
$N=4$ (black cubic), $N=6$ (red circle) and $N=8$ (blue triangle)
respectively. (a): the behavior of $p_{\epsilon}$ for large $d$.
(b): the behavior of $p_{\epsilon}$ for small $d$. It can be seen
that when $d$ is fixed, $p_{\epsilon}$ decreases with $N$ while when
$N$ is fixed, it grows with $d$ to a saturated value
$1-\epsilon^{1/N}$.}
\end{figure}

\begin{figure}
\includegraphics[height=6cm,width=\linewidth]{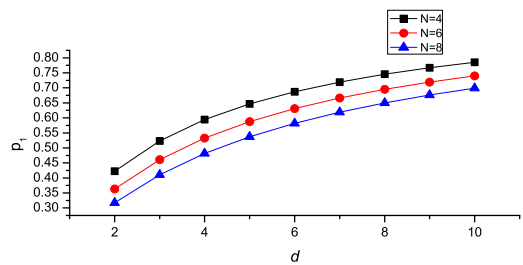}
\caption{\label{fig:3} (color online) For the depolarizing channel,
we show the relation of $p_{1}$ with $d$ and $N$. Here we choose
$\alpha_{i}=1/\sqrt{d}$ for convenience and $N=4$ (black cubic),
$N=6$ (red circle) and $N=8$ (blue triangle) respectively. It can be
seen that when $d$ is fixed, $p_{1}$ decreases with $N$ while when
$N$ is fixed, it grows with $d$.}
\end{figure}

One question arises that for the depolarizing channel, what's the
dynamical behavior of the bipartite entanglement corresponding to
the least balanced partition that vanishes first? As we know, in
qubit case ($d$=2) it vanishes earlier when $N$ grows. Now our
numerical calculation shows its relation with $N$ and $d$ (FIG.3).

One could expect that the dependence of entanglement robustness on
$N$ and $d$ would be similar, since in both cases we are increasing
the dimension of the Hilbert space of each of the bi-partitions. But
in fact their influences on the entanglement robustness are
different. Roughly speaking, entanglement increases with $d$ and
thus it is more robust, while entanglement becomes more fragile with
$N$ since more particle are entangled together and it becomes easier
to be destroyed. We take the depolarizing channel as an example, for
which according to the discussion above, $p_\epsilon$ increases with
$d$ while it decreases with $N$. This can be explained as follows.
If $d$ is fixed, when we increase $N$, we increase the components of
the state. Considering the depolarizing channel acts locally on
every component, the growth of $N$ will make the entanglement more
fragile. In the depolarizing channel $\mathcal
{E}_{1}(\rho)=(1-p)\rho+\frac{p}{d}\mathbf{1}$, $p$ can be regarded
as the probability with which $\rho$ is broken by the channel. The
probability of the $N$-particle state to be broken by the collective
local action of $N$ channels on each particle must grow with $N$
(about $Np$ as a rough estimation), leading that the entanglement
becomes less robust. The probability with which the $N$-particle
state is not broken can be estimated roughly as $(1-p)^{N}$ and this
probability can also be expressed as $\epsilon$ considering the
entanglement decay. Therefore we have $(1-p)^{N}=\epsilon$, leading
to our familiar results. If $N$ is fixed, with the increase of $d$,
the influence of the channel on $\rho$ will be more and more
negligible, which leads the entanglement becomes robust.

{\label{sec:5}} \section{Summary} In this brief report, we mainly
investigate the dynamical property of entanglement of a generalized
$N$-qudit GHZ state under the influence of the depolarizing channel
and the phase damping channel. We study the relation of the
entanglement robustness with $N$ and $d$. First we consider the ESD
time $t_{1}$ and $t_{2}$ respectively of the bipartite entanglement
corresponding to the most balanced partition ($t_{1}$) and the
bipartite entanglement corresponding to the least balanced partition
($t_{2}$). When $d$ is fixed, for the depolarizing channel $t_{1}$
delays when $N$ grows whereas $t_{2}$ becomes earlier. For the phase
damping channel, no ESD happens for any $N$. These results are
qualitatively the same with that of $d=2$. When $N$ is fixed, for
the depolarizing channel both $t_{1}$ and $t_{2}$ grow with $d$
whereas still no ESD happens for the phase damping channel for any
$d$. Next we consider the time at which the bipartite entanglement
becomes arbitrarily small ($t_{3}$). When $d$ is fixed the scaling
relation between $p_{\epsilon}$ and $N$ is totally independent of
$d$ therefore the same with the result in Ref.\cite{r4} for both
channels. When $N$ is fixed, for the depolarizing channel $t_{3}$
grows with $d$ whereas it's independent of $d$ for the phase damping
channel. There are many other multiqudit entangled states such as
generalized $N$-qudit W-state \cite{r6} and other quantum channels.
It is not clear how the behaviors of the entanglement differ in
those situations. This is worth being studied further.

{\label{sec:level6}} \section{Acknowledgements} H.F. acknowledges
the support by "Bairen" program, NSFC grant (10674162) and "973"
program (2006CB921107).

\newpage
\bibliography{apssamp}

\end{document}